\documentclass[10pt,aps,prb,twocolumn,superscriptaddress,preprintnumbers,amsmath,amssymb,floatfix]{revtex4-1}
\usepackage[english]{babel}
\usepackage{graphicx}
\usepackage{amsmath}
\usepackage{amssymb}
\usepackage{dcolumn}
\usepackage{times}
\usepackage{psfrag}

\renewcommand{\AA}{\text{\r{A}}}

\begin{document}

\title{Ternary Semiconductors NiZrSn and CoZrBi with half-Heusler structure: a first-principles study}
\author
{Gregor Fiedler and Peter Kratzer}
\affiliation{Fakult\"at f\"ur Physik, Universit\"at Duisburg-Essen, Campus Duisburg, 
Lotharstr.~1, 47057 Duisburg, Germany}
\date{\today}
\begin{abstract}
The ternary semiconductors NiZrSn and CoZrBi with $C1_b$ crystal structure are introduced by calculating their basic structural, electronic and phononic properties using density functional theory. 
Both the gradient-corrected PBE functional and the hybrid functional HSE06 are employed. While NiZrSn is found to be a small-band-gap semiconductor (E$_g = 0.46$ eV in PBE and 0.60 eV in HSE06), CoZrBi has a band gap of 1.01 eV in PBE (1.34 eV in HSE06). Moreover, effective masses and deformation potentials are reported. 
In both materials $ABC$, the intrinsic point defects introduced by species $A$ (Ni or Co) are calculated. 
The Co-induced defects in CoZrBi are found to have a higher formation energy compared to Ni-induced defects in NiZrSn. 
The interstitial Ni atom (Ni$_i$ ) as well as the V$_{\rm Ni}$Ni$_i$ complex introduce defect states in the band gap, 
whereas the Ni vacancy (V$_{\rm Ni}$) only reduces the size of the band gap. 
While Ni$_i$ is electrically active and may act as a donor, the other two types of defects may compensate extrinsic doping. 
In CoZrBi, only the V$_{\rm Co}$Co$_i$ complex introduces a defect state in the band gap. 
Motivated by the reported use of NiZrSn for thermoelectric applications,  the Seebeck coefficient of both materials, both in the p-type and the n-type regime, is calculated. 
We find that CoZrBi displays a rather large thermopower of up to 500$\mu$V/K when p-doped, whereas NiZrSn possesses its maximum thermopower in the n-type regime. 
The reported difficulties in achieving p-type doping in NiZrSn could be rationalized by the unintended formation of Ni$_i^{2+}$ in conjunction with extrinsic acceptors, resulting in their compensation. Moreover,  
it is found that all types of defects considered, when present in concentrations as large as 3\%,  tend to reduce the thermopower compared to ideal bulk crystals at $T=600$~K. 
For NiZrSn, the calculated thermodynamic data suggest that additional Ni impurities could be removed by annealing, leading to precipitation of a metallic Ni$_2$ZrSn phase. 
\end{abstract}

\maketitle
\section{Introduction}
While semiconductor technology has mostly used elementary semiconductors such as Si or Ge, or binary compounds\footnote{In this article,  alloys with binary lattice, such as AlGaAs or InGaN, count as variants of binary materials.}, such as GaAs or InP, semiconductors crystallizing in more complex structures with three inequivalent lattice sites are much less explored. 
Yet, true ternary materials could offer significant advantages over binary materials. One important class of materials $ABC$, having the  $C1_b$ structure  (space group $F \bar 4 3m$) consists of  a main group element $C$ from column IV or V of the periodic table, while $B$ stands for an early transition metal, and $A$ is a late transition metal. These are sometimes referred to as half-Heusler compounds, because they are related to the closed-packed Heusler alloys with formula $A_2 BC$ by leaving every second $A$ site empty. In this paper, we will prefer to describe the crystal structure as a rocksalt structure of the $BC$ compound with  exactly  {\em each second interstitial} site of the rocksalt structure being occupied by an atom of species $A$. 
Alternatively, these compounds are sometimes described as filled tetrahedral structures derived from the zincblende lattice $AC$ where half of the tetrahedral interstitial sites are occupied by $B$. 
Electron counting arguments~\cite{Graf:11} suggest that the compound may be semiconducting if it contains overall 18 valence electrons. 
Since, given this restriction,  $A$, $B$ and $C$ can still be varied over some range of the periodic table, the resulting compounds may display a large variety of properties, both structural (e.g. lattice parameters, elastic moduli) as well as electronic (e.g. band gap)\cite{Yang:08}. 
Moreover, heterostructures of  two such ternary materials, e.g. $ABC$ and $A'BC'$, may allow crystal growers to fabricate materials that are very similar in two or more properties (e.g. lattice constant {\em and} valence band position), while differing significantly in another property. Such a selection of several properties by tuning of the chemical composition is rarely possible in heterostructures of binary compounds, but becomes feasible due to the enhanced flexibility of ternary materials. One example of rational materials design is the theoretical prediction~\cite{Zakutayev:13} of the ternary semiconductor CoTaSn as an optically transparent material that allows for p-type conductivity when doped. Thermoelectric materials provide further examples of materials by design, where the thermal conductivity of half-Heusler alloys has been addressed recently \cite{Carrete:14x}, albeit other recent computational design studies have focussed on doped binary thermoelectric materials~\cite{Opahle:12,Bera:14}. 

While some screening of the ternary compound space for basic properties has been carried out by high-throughput computational methods~\cite{Zhang:12,Carrete:14x,Yang:08}, a detailed description of materials properties that are essential for semiconductor technology, e.g. effective masses and deformation potentials of the charge carriers, is still elusive for most ternary $C1_b$ compounds. 
In this  article, we aim at a comprehensive characterization (from a theoretical perspective) of two representative compounds, NiZrSn and CoZrBi. 
The interest in these materials is mostly driven by thermoelectrics. 
In particular, theoretical studies of NiZrSn with focus on thermoelectric applications have been published recently.\cite{Page:15,Fecher:16}  
From an experimental perspective, half-Heusler materials with similar composition, e.g. NiTiSn or Ni(Zr,Hf)Sn or CoTiSb,  have already shown promising performance. 
Empirically, it is observed that materials $ABC$ with species $C$ taken from group IV 
are convenient to use as n-type semiconductors~\cite{Uher:99,Ouardi:10}, while those with $C$ from group V are conveniently used in combination with p-type doping. 

In addition to electronic and phononic band structure data, we will present in this paper a theoretical study of the role of intrinsic defects in NiZrSn and CoZrBi. Finally, the Seebeck coefficient of both pure and defect-containing samples will be calculated and compared. 
Our results 
allow us to explain why NiTiSn and NiZrSn are  often reported to be intrinsically n-type~\cite{Uher:99,Ouardi:10}, and why p-type conductivity is difficult to achieve even if a group-III element with concentrations of several percent is added~\cite{Ouardi:10}. 
According to our findings, interstitial Ni atoms may spontaneously be created during sample growth which introduce defect states near the conduction band edge and thus counteract any intended p-type conductivity. 
Starting from a material with a group-V-element on lattice site $C$ in conjunction with Co on lattice site $A$ seem to be more promising for achieving p-type conductivity, since this allows one to introduce hole carriers by substituting the group-V-element by a group-IV atom.  An examples is Zr$_{0.5}$Hf$_{0.5}$CoSb$_{0.8}$Sn$_{0.2}$ studied in Ref.~\onlinecite{Yan:10,Rausch:14}. Since the formation energy of Co interstitials is found to be very high, the risk of introducing unwanted compensating donors is minimized in this material. 
Since the lattice constants of NiZrSn and CoZrBi differ by less than 1.5\%, these two materials lend themselves to the fabrication of lattice-matched heterostructures. The advantages of such heterostructures for thermoelectric applications have been analyzed recently from a theoretical perspective \cite{Fiedler:15b}. They could pave the way to compound materials with a higher thermoelectric figure of merit, as a significant reduction of the thermal conductivity perpendicular to the layers  has been demonstrated both theoretically and experimentally\cite{Jaeger:14,Jakob:15,Komar:16}.  

\section{Computational methodology and tests}
All properties in the present study have been obtained from first principles, using density functional theory. 
As standard approach, the generalized gradient approximation (PBE functional~\cite{PeBu96}), for the exchange-correlation energy of the electrons is employed. We work with both an all-electron description and a pseudopotential description. The latter proves to be more efficient for larger systems, including point defects. 

Specifically, the pseudopotential calculations in conjunction with a plane-wave basis for the wavefunctions have been carried out with the software package 
Quantum Espresso~\cite{PWSCF}. For Zr, the occupied $3s$ and $3p$ shells have been treated as valence electrons; i.e., the pseudopotential is constructed from the configuration
$3s^2$ $3p^6$ $4s^2$ $3d^2$. For Co and Ni, the $3d$ and $4s$ electrons were treated as valence electrons, while for Sn, the $5s^2$ $5p^2$ configuration with four valence electrons was used. All these elements were treated by ultra soft pseudopotentials (USPPs), as obtainable from the Quantum Espresso web-site~\cite{QE-website}. 
Only for Bi that has anionic character in CoZrBi, a norm-conserving pseudopotential with five valence electrons was used for convenience. 
The validity of the pseudopotential results was checked by comparing to all-electron calculations (see below). 
The excellent agreement found justifies the use of pseudopotentials, and also demonstrates the compatibility of ultra-soft and norm-conserving pseudopotentials used in this work. 
With these settings, we obtained converged results by using a cut-off energy of 45~Ry for the wavefunctions, while a cut-off energy of 450~Ry was used to represent the charge density.

For the all-electron approach, we used the software FHI-aims~\cite{FHIaims} that represents the wavefunctions by a linear superposition of atomic orbitals numerically defined on a real-space grid. We use this code mostly for hybrid functional calculations with the HSE06~\cite{Heyd2004} functional, but test calculations were also performed with PBE to verify the pseudopotential results. The FHI-aims code uses the very efficient  resolution-of-the-identity method~\cite{FHIaimsHSE} to represent the Fock operator within the orbital basis. For many binary semiconductors the generalized Kohn-Sham approach within HSE was found to give more realistic band structures~\cite{PaMa06}, while at the same time being able to describe structural properties of the materials.

First, we obtain optimized lattice constants for both NiZrSn and CoZrBi, using both the PBE and HSE06 functionals. The results are reported in Table~\ref{tab:lattice_const} and compared to experiment.
We find that the PBE pseudopotential calculations overestimate the lattice constant  by 1\% and 0.7\%, respectively, 
while the HSE06 calculations slightly underestimate the lattice constants.

Moreover, the elastic properties of both NiZrSn and CoZrBi have been calculated (Table~\ref{tab:effMass}). 
Since both materials crystallize in the cubic system, three elastic constants, $C_{11}$, $C_{12}$ and $C_{44}$ are sufficient to describe the elastic properties completely. 
The procedure for extracting elastic constants from the calculations is described in detail in the Appendix. 
Both materials are characterized by a large bulk modulus, but relatively small resistance to shear deformations, evidenced by the small values of $C_{12}$ and $C_{44}$ as compared to $C_{11}$. 
Both the PBE and the HSE06 functional give a comparable description of the elastic properties. While $C_{11}$ is only about 10\% to 15\% enhanced in HSE06 compared to PBE, the stiffness against shear deformations comes out clearly higher when the HSE06 functional is used. 

\begin{table}[tbh]
%	\centering
		\begin{tabular}{l|r|r|r}
		\, & PBE  & HSE06 &  experiment \\
		\hline
		CoZrBi & 6.25 &  6.15 & 6.19 \\
		NiZrSn & 6.15 & 6.10  & 6.11 \\
		\end{tabular}
		\caption{Lattice constants in {\AA} for CoZrBi und NiZrSn from DFT calculations using the PBE and the HSE06 functional, respectively. The experimental values are taken from Ref.~\onlinecite{Evers:97}  and \onlinecite{Pearson:91}, respectively.}
		\label{tab:lattice_const}
\end{table}

\begin{figure}[bth]
		\includegraphics[width=0.45\textwidth]{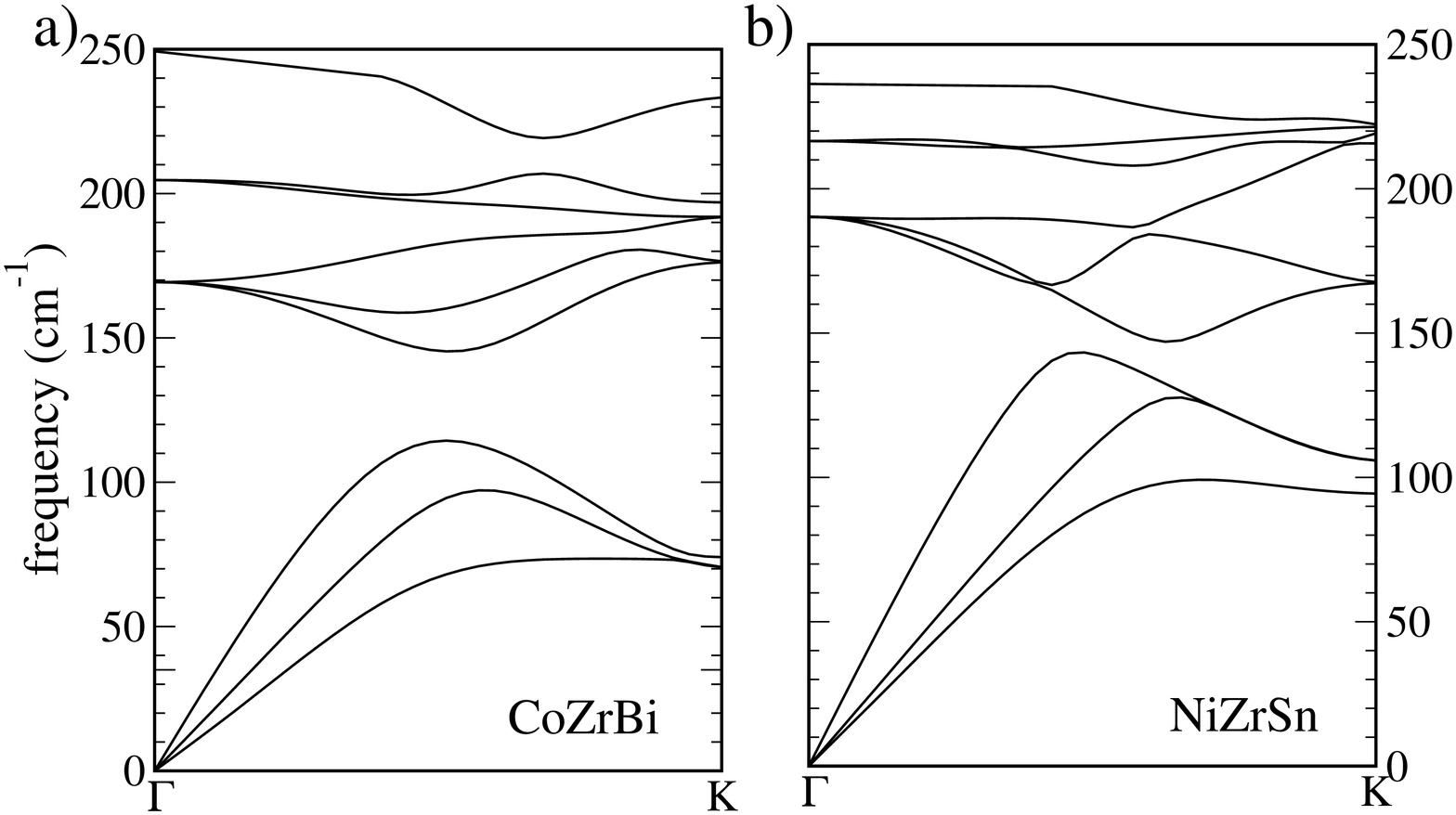}
\caption{Phonon spectra of (a) CoZrBi  and (b)  NiZrSn  in (110) direction.}
\label{fig:phonons}
\end{figure}

In order to probe the stability of the structures against small perturbations, phonon spectra were calculated using 
density functional perturbation theory\cite{BaGi01}, as implemented in Quantum Espresso. 
In Fig.~\ref{fig:phonons}, calculated phonon spectra are shown along the (110) direction. 
The calculations were performed on a $4 \times 4 \times 4$ $\textbf{q}$ mesh in the phonon Brillouin zone, and force constants in real space derived from this input are used to interpolate between {\bf q}-points and to obtain the continuous branches of the phonon band structure. 
All nine modes are found to have real frequencies, i.e., the half-Heusler crystals are stable with respect to small distortions of the lattice. 
The plot along  (110) direction is instructive, because in this direction the degeneracy of the modes is completely lifted inside the Brillouin zone for any wave vector $\textbf{q} \ne 0$.  

Both NiZrSn and CoZrBi show six optical modes in the energy range of 20~meV to 30~meV (160~cm$^{-1}$ to 250~cm$^{-1}$). 
The  displacements characteristic of these modes have their leading contributions from the Zr atoms, which are displaced relative to the Ni and Co atoms. The similar masses of Ni and Co, in conjunction with the similar bond strengths in both materials, result in quite similar optical modes. 
The three acoustic modes in NiZrSn and CoZrBi are qualitatively similar, but quantitatively different. 
A common feature is a maximum of the dispersion at intermediate $\textbf{q}$ values, which indicates the importance of second-nearest neighbor and even more long-ranged interactions. 
Mostly due to the larger mass density of CoZrBi, the acoustic modes are 
considerably softer than in NiZrSn. 
The speed of sound, as determined  from the slope of the phonon dispersion, is 5220~m/s, while the transverse modes propagate at a speed of 3060~m/s and 2610~m/s in NiZrSn. The corresponding values for CoZrBi are  4355~m/s, 2727~m/s and 1931~m/s, i.e., about 20\% lower. 
We note that similar values (within an error of 10\%) for the speed of sound can be obtained simply from the elastic constants (Table~\ref{tab:effMass}) and the mass densities of the two  materials. 
Since the acoustic phonon branches in CoZrBi are soft, their dispersion curves remain clearly separated from the optical phonon branches, and a gap in the phonon spectrum occurs at 130 to 140 cm$^{-1}$. This is different from NiZrSn, where overlap of acoustic and optical branches is observed for intermediate to large $\textbf{q}$ values.

\section{Results}
\subsection{Electronic density of states and band structure \label{sec:bands}}

It is highly remarkable that the compounds NiZrSn and CoZrBi are semiconductors, while all constituent elements are metallic. 
In fact, the band gap in these materials must arise as a hybridization gap between the constituents' $d$ orbitals split due to crystal symmetry. 
As described in Ref.~\onlinecite{Ogut:95}, the stability of the $C1_b$ structure is mainly due to the $B2$ (rocksalt) lattice formed by the early transition metal (Zr in our case) and the group-IV or group-V species. 
The cubic environment of Zr in the $B2$ structure splits its $4d$ orbitals into subgroups belonging to the  three-fold $t_{2g}$ and two-fold $e_g$ representation.  
However, the band complexes derived from these subgroups fill overlapping energy intervals, thus rendering ZrSn and ZrBi in the rocksalt structure metallic at their respective equilibrium lattice constants. 
Inserting Ni or Co into  each second interstitial site widens the lattice by about 5\%, which reduces the band overlap. This, together with the symmetry reduction, is responsible for the opening of a band gap.

\begin{table}[tbh]
\begin{tabular}{l|c|c|c|c}
 & \multicolumn{2}{c|}{NiZrSn} & \multicolumn{2}{c}{CoZrBi} \\
\hline
 functional & PBE & HSE06 & PBE & HSE06 \\
 \hline
 $C_{11}$ & 234 & 251 & 227 & 264 \\
 $C_{12}$ & 58 & 79 & 54 & 72 \\
 $C_{44}$ & 64 & 72 & 41  & 58 \\
 \hline
 $E_{\rm gap}$ & 0.46 & 0.60 & 1.01 & 1.34 \\
 $a_v$ & 5.0 & 5.2 & 4.7 & 4.9 \\
 $a_c$ & 4.0 & 4.3 & 3.2 & 3.8 \\ 
 \hline
 $\Xi_u$  & 0.80 & 0.80 & 0.88 & 0.80 \\
 $m_{||}$ &3.45 & 2.73 & 8.63 & 4.86 \\
 $m_{\perp}$ & 0.39 & 0.31 & 1.20 & 0.79 \\
 \hline
 $b$ & $-1.72$ & $-1.92$ & $-1.50$ & $-1.92$ \\
 $d$ & $-4.27$ & $-4.54$ & $-4.14$ & $-4.49$ \\
 $m_{lh,\Gamma L}$ & 0.26 & 0.46 & 0.19 & 0.39 \\
 $m_{lh, \Gamma X}$ & 0.41 & 0.36 & 0.29 & 0.30 \\
 $m_{hh, \Gamma L}$ & 2.35 & 3.26 & 1.36 & 2.21 \\
 $m_{hh, \Gamma X}$ & 0.88 & 0.75 & 0.65 & 0.56 \\ 
\end{tabular}
\caption{Elastic constants in GPa, band gaps and deformation potentials in eV, and effective masses in units of the free electron mass, 
for NiZrSn and CoZrBi, calculated with both the PBE and the HSE06 density functional.}
\label{tab:effMass}
\end{table}

\begin{figure}[tbh]
%	\centering{
		\includegraphics[width=0.42\textwidth,clip]{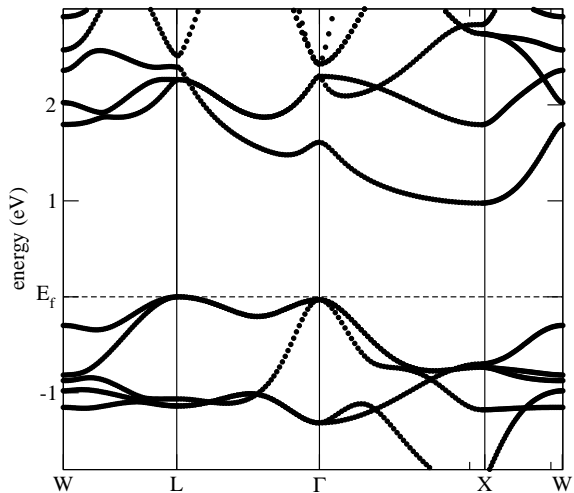}
		\caption{Kohn-Sham band structure of CoZrBi}
		\label{fig:bands-ZrCoBi}
\end{figure}

\begin{figure}[tbh]
%	\centering{
		\includegraphics[width=0.42\textwidth,clip]{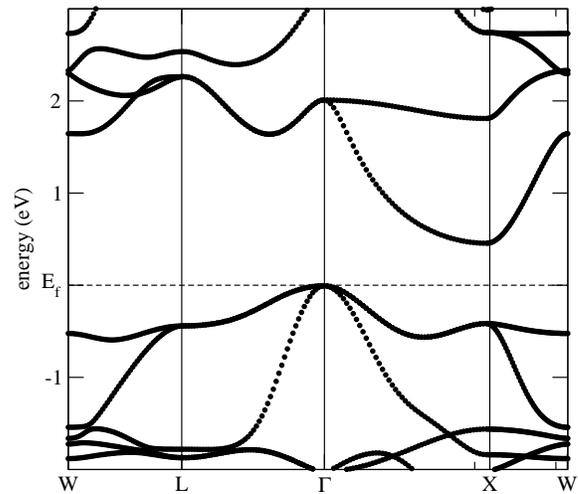}
		\caption{Kohn-Sham band structure of NiZrSn}
		\label{fig:bands-ZrNiSn}
\end{figure}

We employ two different functionals for the electronic structure calculations: 
The PBE functional was used to calculate Kohn-Sham bands both at the all-electron level (FHI-aims code) and in the pseudopotential approximation (Quantum Espresso code). Excellent agreement between both approaches is found. 
The band structures displayed in Figs.~\ref{fig:bands-ZrCoBi} and \ref{fig:bands-ZrNiSn} were calculated at the theoretical equilibrium lattice constants, reported in Table~\ref{tab:lattice_const}, with Quantum Espresso. 
The Kohn-Sham band gap in NiZrSn is found to be an indirect gap $\Gamma - X$ with a size of 0.46~eV. 
This value is in good agreement with previous calculations~\cite{Zou:13,Do:14}, but much larger than reported typical experimental values of 0.18~eV\cite{Aliev:90} and 0.13~eV~\cite{Schmitt:15}.  
In CoZrBi, the band gap is indirect as well ($L - X$ in the PBE approximation, $\Gamma - X$ in HSE06) and has a size of 1.01~eV. Here, experimental values are not available for comparison. 

The calculations were repeated for the hybrid functional HSE06, using an admixture of $\alpha=25$\% exact exchange and a screening length of $\omega=0.11$~bohr$^{-1}$.  
The values obtained from the all-electron calculations at the HSE06 equilibrium lattice constants are 0.60~eV and 1.34~eV, respectively.
Unlike for polar materials with a charge-transfer gap, 
where HSE06 typically increases the gap by a factor two, 
the enlargement of the band gaps in the half-Heusler compounds by using HSE06 are rather modest (about 30\% increase). This seems to indicate that the orbitals forming the conduction band and the valence band, respectively, are about equally affected by the admixture of exact exchange.
Most importantly, the hybrid functional calculations indicate that the true quasiparticle band gap of ideal half-Heusler semiconductors is larger than the Kohn-Sham gap, as is the case in most semiconductors studied so far. The discrepancy between the apparently very small band gap in experiment and the calculated values most likely originates from a state induced in the gap due to defects (see Section~\ref{sec:defects}). 

For controlled application of half-Heusler materials in future semiconductor technology, knowledge of the parameters governing electronic transport, such as band effective masses and deformation potentials, is essential. In Table~\ref{tab:effMass} we have collected our results. 
The effective masses were calculated from the second derivatives of the bands around the $\Gamma$ point for the valence band, and around the $X$ point for the conduction band. 
Neglecting spin-orbit interaction, the valence band maximum is triply degenerate, with two light-hole bands and one heavy-hole band, as in most cubic semiconductors. 
The effective masses of the hole differ along the $\Delta$ line ($\Gamma X$ direction) and the $\Lambda$ line ($\Gamma L$ direction) in the Brillouin zone. 
The relatively flat heavy-hole band in $\Gamma L$ direction gives rise to an effective mass (quoted in multiples of the bare electron mass) much larger than unity. 
The effective masses along $\Lambda$ calculated with the HSE06 functional are generally somewhat higher than their PBE counterparts.
The conduction band minima (CBM) have the shape of six pockets located at the Brillouin zone edges. The principal axes of the effective mass tensor coincide with the coordinate axis of $\mathbf{k}$-space. In each pocket, the motion of the electrons is characterized by a heavy electron mass along $\Delta$, and identical light electron masses in the two perpendicular directions.  
As a consequence, the charge carriers in lightly doped n-type material are expected to be highly mobile along the directions of low effective mass, whereas the density of states at the conduction band edge is still high due to the large transverse effective mass and the six-fold degeneracy. The latter finding is responsible for the relative  insensitivity of the Fermi level position with respect to the carrier concentration in n-doped half-Heusler materials.  
The combination of both features makes the n-type material very attractive for thermoelectric applications. 
The effective masses for NiZrSn obtained in PBE are in excellent agreement with recent work using all-electron DFT calculations~\cite{Fecher:16}. 
Using the HSE06 functional, the effective electron masses are clearly reduced with respect to the PBE values, indicating that the HSE06 conduction band is somewhat more dispersive than the PBE conduction band. 

In both NiZrSn and CoZrBi, hydrostatic pressure shifts the valence band and conduction band edges to lower energies, This is reflected in the positive values of the deformation potentials $a_v$ and $a_c$ reported in Tab.~\ref{tab:effMass}. Since $a_v > a_c$, the band gap widens upon the application of hydrostatic pressure.
Details of the method of calculating the deformation potentials are given in the Appendix. 
The behavior of the conduction band valleys at the $X$ point is characterized by the two deformation potentials $\Xi_d$ and $\Xi_u$. The latter describes the response to uniaxial strain and is important for the scattering of carriers by acoustic phonons. 
The values found for $\Xi_u$ are about an order of magnitude smaller than for the conduction band of silicon, which again points to the high mobility that could be achieved in lightly doped n-type samples.
The behavior of the valence band edge under strain is described by the deformation potentials $a_v$, $b$ and $d$. While $b$ describes the splitting of the degenerate valence band maximum under a tetragonal distortion, the deformation potential $d$ describes the splitting under a shear deformation along the (111) crystallographic axis.
Both $b$ and $d$ are found to take on negative values for both materials. This is the same qualitative behavior as observed for polar binary materials, e.g. GaAs and InAs.

\begin{figure}[tbh]
		\includegraphics[width=0.45\textwidth]{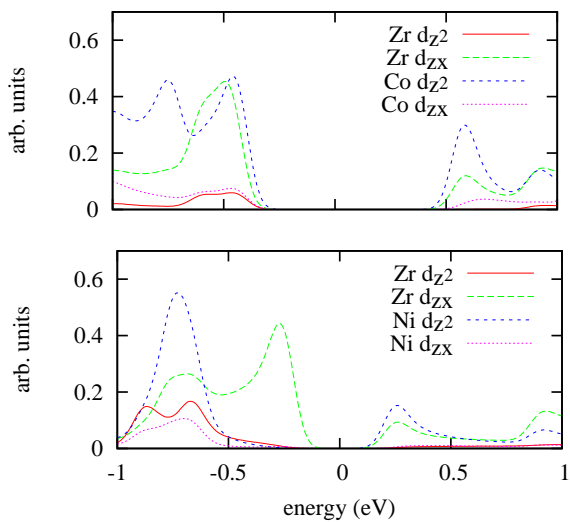}
\caption{Projected DOS of the $e_g$ and $t_{2g}$ $d$ orbitals of Zr and Co in CoZrBi (upper panel) and Zr and Ni in NiZrSn (lower panel) around the band gap.}
\label{fig:DOS-comparison}
\end{figure}

In the following, we give an interpretation of the calculated band structures in terms of atomic orbitals on the basis of the orbital-projected density of states (Fig.~\ref{fig:DOS-comparison}).
The valence bands are formed by hybridization between Zr $4d$ orbitals  of $e_{g}$ symmetry ($d_{z^2}$ and $d_{x^2-y^2}$) with the $5p$ orbitals of Sn and the $6p$ orbitals of Bi, respectively, with the Zr $4d$ orbitals dominating the valence band character near the $\Gamma$ point.  
Filling each second vacancy in the rocksalt structure of ZrSn or ZrBi by the Ni or Co atoms lowers the overall symmetry from $O_h$ to $T_d$. 
This symmetry break is essential for the symmetry character of the lowest conduction band. It is formed from the hybridization of unoccupied Zr $4d$ orbitals of $t_{2g}$ symmetry with the $3d$ orbitals of Ni or Co.
This conduction band reaches its minimum at the $X$ point, where the $\mathbf{k}=(100)$ vector points in the direction along the Zr--Ni--Zr--Ni or Zr--Co--Zr--Co chains. 
As can be seen from Fig.~\ref{fig:DOS-comparison}, the $3d$ states of Ni (which are also split into $t_{2g}$ and $e_g$ manifolds) are energetically lower than the Co $3d$ orbitals. Therefore, the Ni $3d$ states hybridize efficiently with the unoccupied Zr $4d$ states; and both contribute equally to the conduction band (Fig.~\ref{fig:DOS-comparison}b). As a result of the strong hybridization, the conduction band is highly dispersive; this leads to a relatively small band gap in NiZrSn as compared to CoZrBi (see Fig.~\ref{fig:bands-ZrCoBi} and  \ref{fig:bands-ZrNiSn}). On the other hand, the lower-lying Co $3d$ states contribute more to the valence bands in CoZrBi, resulting in about equal contributions of Zr and Co at the valence band maximum (Fig.~\ref{fig:DOS-comparison}a).  
Moreover, the hybridization of Zr and Co $d$ orbitals along the Zr--Co--Zr--Co chains in $(111)$ direction gives rise to states at the $L$ point of the Brillouin zone. When using the PBE functional, the near-degeneracy of Zr $4d$ and Co $3d$ orbitals results in strong hybridization that causes the $L$ point to become the VBM. 
In a band structure obtained with the HSE06 functional, however, the somewhat different energetic alignment of these bands causes the VBM to occur at the $\Gamma$ point. 
In summary, due to the lower-lying $3d$ orbitals of Co, their hybridization with Zr $4d$ orbitals is more difficult (compared to the hybridization between Ni $3d$ and Zr $4d$ orbitals). For this reason, the conduction band in CoZrBi is less dispersive than in NiZrSn, resulting in a larger band gap in CoZrBi. 

Moreover, our orbital analysis indicates that the orbitals of the third constituent, Sn or Bi, contribute little to the band edges. Therefore, the effects of spin-orbital coupling, which one could expect to be sizable for a heavy species such as Bi, have little effect on the structure of the band edges.  
Supplementary calculations using the WIEN2k code~\cite{ScBl03}, treating spin-orbit coupling as a second-order variation to the unpolarized wavefunction, gave values of 
$\Delta E = 0.016$~eV and  
$\Delta E = 0.148$~eV for the spin-orbit splitting of the triply degenerate VBM at $\Gamma$ for NiZrSn and CoZrBi, respectively.

\subsection{Defects of the $A$ sublattice \label{sec:defects}}

In this Section, we concentrate on the defects due to the $A$ atoms (Ni or Co atoms, in our case) in the half-Heusler compound $ABC$. 
This is because the two other species, $B$ and $C$, form a rather stable rocksalt lattice structure $BC$ with partial charge transfer between these two species, and defect formation energies in the $BC$ lattice therefore tend to be very high. 
For the application as thermoelectric materials, the defects giving rise to in-gap states are the most important ones.\cite{Miyamoto:08} 
From the orbital analysis in the Section~\ref{sec:bands}, it is obvious that it is species $A$ that has the potential of inducing such in-gap states, since the atomic orbitals of $A$ (of Ni and Co) contribute {\em both} to the valence {\em and} the conduction band states. 
Qiu {\em et al.}, Ref.~\onlinecite{Qiu:10}, performed calculations for defects resulting from swaps of Zr and Ni atoms.  They observed a narrowing of the band gap, but no states in the gap. In addition, we considered the displacement of a Zr atom into a vacant interstitial site but noticed that the atomic configuration relaxed back to the ideal lattice positions. 
For these reasons, we limit the discussion below to $A$-related defects.

For the thermoelectric applications, the half-Heusler materials are sometimes deliberately synthesized \cite{Romanka:13,Zou+Weidenkaff} with an excess of element $A$ which is supposed to occupy each second interstitial site in the rocksalt lattice $BC$. 
While the solubility of the $A$ species in the half-Heusler alloy is rather limited \cite{Page:15},  concentrations of $A$ of the order of a few percent may still be reached by occupying additional interstitial sites by $A$ atoms. The thermodynamics for this process can be estimated by calculating the stability of the (full) Heusler alloy $A_2BC$ which is ultimately formed after {\em all} interstitial sites are filled with $A$. 
Our PBE calculations show that, for NiZrSn, the formation of Ni$_2$ZrSn from NiZrSn and elemental Ni is slightly exothermic by 0.10~eV. This can be seen as an indication that the formation energy of additional Ni atoms in the lattice is not too high. For CoZrBi, the formation of the full Heusler alloy is found to be endothermic even in the presence of elemental Co. Hence, for this material, one would also expect a high formation energy for additional Co atoms in the half-Heusler lattice.

We note that for defect concentrations at the percent level, as often found in thermoelectric materials, the position of the Fermi level for the defective material is controlled by the defect level or defect bands themselves. To model this situation, we use a $2 \times 2 \times 2$ supercell (96 atoms in total) with one additional Ni or Co atom. This corresponds to an off-stoichiometric composition $A_{1+x}BC$ with $x=0.03$. Taking this concentration as representative for the material used in real applications, we proceed by calculating band structures and thermoelectric coefficients for this supercell (see Section~\ref{sec:Seebeck} below). 
For the supercell calculations, a $4 \times 4 \times 4$ {\bf k}-point mesh has been used to sample the Brillouin zone. 
The atomic positions were relaxed in the PBE calculations. 
For the HSE06 calculations, the geometries relaxed with PBE were used as input after scaling them to the HSE06 lattice constant. 
  and held fixed. 
In case of very low defect concentrations, as can be achieved in traditional semiconductors such as Si or GaAs, one usually discusses the formation energy of a defect as a function of the position of the Fermi level $E_F$ in the band gap, relative to the valence band edge $E_{\rm VB}$, so as to model charge transfer from the defect under investigation to possible other (intentionally or unintentionally added) 
impurities of the sample. 
In order to be able to make contact with these types of studies, we also calculated the formation energy for charged defects, using the method of a charged supercell with a fixed, homogeneously distributed compensating background charge.\cite{VandeWalle2004} 
The formation energy of the defect is given by 
\begin{equation}
E_{\rm form}(\mu_A, E_F) = E_{\rm tot}- E_0 - \Delta n_A \mu_A + q (E_F + E_{\rm VB})
\end{equation}
Here $E_{\rm tot}$ is the total energy of the supercell with the defect, $E_0$ is the energy of an ideal, unperturbed supercell, and $\mu_A$ is the chemical potential of species $A$, $q$ represents the charge state, and $\Delta n_A$ denotes the deviation of the stoichiometry of species $A$ between the cell with and without the defect. 

\begin{figure}[h]
\includegraphics[width=0.45\textwidth]{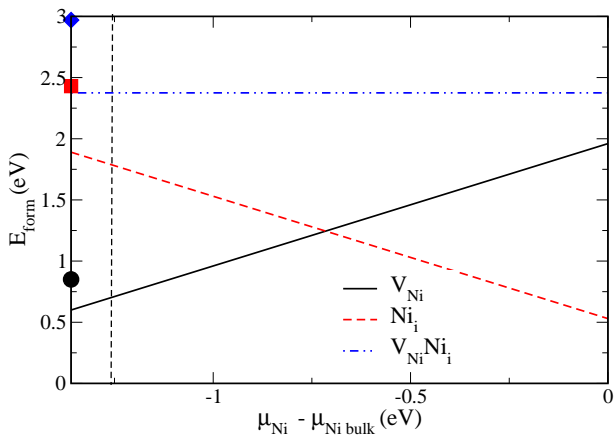}
\caption{Formation energies of neutral Ni defects in dependence on the chemical potential of Ni. While the lines are calculated with the PBE functional, the symbols on the left axis mark the results with the HSE06 functional. The leftmost value of $\mu_{\rm Ni}$ corresponds to chemical equilibrium with NiZrSn and ZrSn. The dashed vertical line corresponds to chemical equilibrium with Ni$_2$ZrSn. Only the narrow stripe to the left of width 0.1~eV is thermodynamically accessible.}
\label{fig:formZrNiSn}
\end{figure}

First we discuss the case of Ni-related defects in NiZrSn. 
In addition to the additional Ni atoms on interstitial sites (Ni$_{i}$, $\Delta n_{\rm Ni} = +1$ ), we also consider Ni vacancies (V$_{\rm Ni}$, $\Delta n_{\rm Ni} = -1$) and antistructure defects (Ni$_i$V$_{\rm Ni}$, $\Delta n_{\rm Ni} = 0$) where the total number of Ni atoms is conserved, but a single Ni atom is displaced from its ideal lattice site in the $C_{1b}$ structure by a vector $\vec d \parallel (110)$ to a vacant neighboring interstitial position.
Figure~\ref{fig:formZrNiSn} shows the formation energy for neutral defects of these three types as  function of the chemical potential of nickel, $\mu_{\rm Ni}$. 
The thermodynamically allowed range of $\mu_{\rm Ni}$ is bounded to the left by its value in 
NiZrSn, and bounded to the right by its value in Ni$_2$ZrSn, 
The size of the admissible chemical potential interval is given by the enthalpy of formation of Ni$_2$ZrSn from NiZrSn and Ni, which amounts to 0.10 eV (see above) in the PBE calculations. 
As seen from Fig.~\ref{fig:formZrNiSn}, the antistructure defect is highest in energy for all admissible values of $\mu_{\rm Ni}$, followed by  the Ni vacancy and the interstitial Ni atom. 
The latter two have smallest formation energy at the left and the right boundary of the chemical potential interval, respectively, taking values of 0.60~eV and 1.79~eV in the PBE functional. Only in the extreme case of equilibrium with elemental Ni, the formation energy of Ni$_i$ drops to 0.56~eV (right edge of Fig.~\ref{fig:formZrNiSn}) Using the HSE06 functional, somewhat higher values are obtained, see Tab.~\ref{tab:Eform} for comparison.

\begin{table}[tbh]
\begin{tabular}{l|l|r|r|r}
$A$ in $ABC$   & functional  &  V$_A$  & $A_i$  & V$_A A_i$ \\
\hline 
Ni in NiZrSn  & PBE  &   0.60  &  1.89 & 2.38 \\
Ni in NiZrSn  & HSE06  & 0.85 & 2.43 &  2.91 \\ 
\hline
Co in CoZrBi  & PBE  &   1.41 & 1.02 & 2.97 \\ 
\end{tabular}
\caption{Formation energy in eV of the electrically neutral defects of the $A$ sublattice in NiZrSn and CoZrBi. The values for NiZrSn are given for chemical equilibrium with NiZrSn and ZrSn (Ni-deficient conditions), while the values for CoZrBi are given in equilibrium with bulk Co (Co-rich conditions).
\label{tab:Eform}} 
\end{table}

\begin{figure}[tbh]
\includegraphics[width=0.47\textwidth,clip]{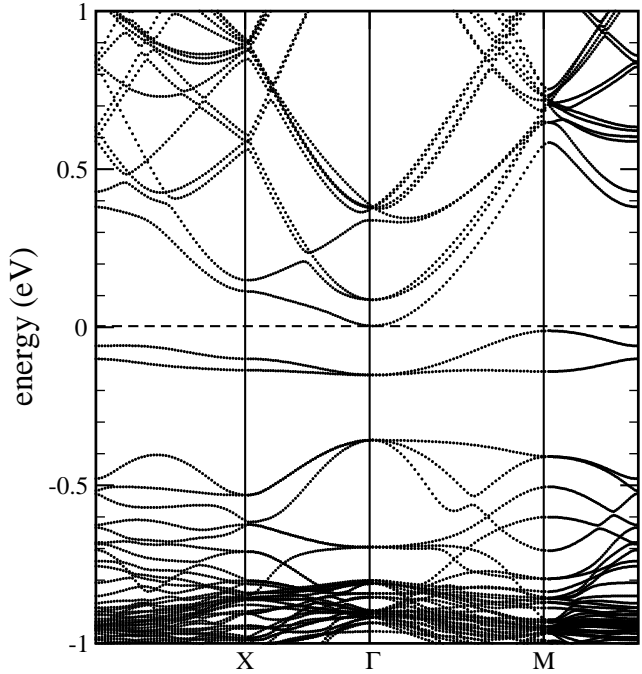}
\caption{Dispersion of the defect state caused by Ni$_i$ in a supercell with 3\% defect concentration.}
\label{fig:bandsZrNiSn+Ni}
\end{figure}

\begin{figure}[bh]
\includegraphics[width=0.47\textwidth,clip]{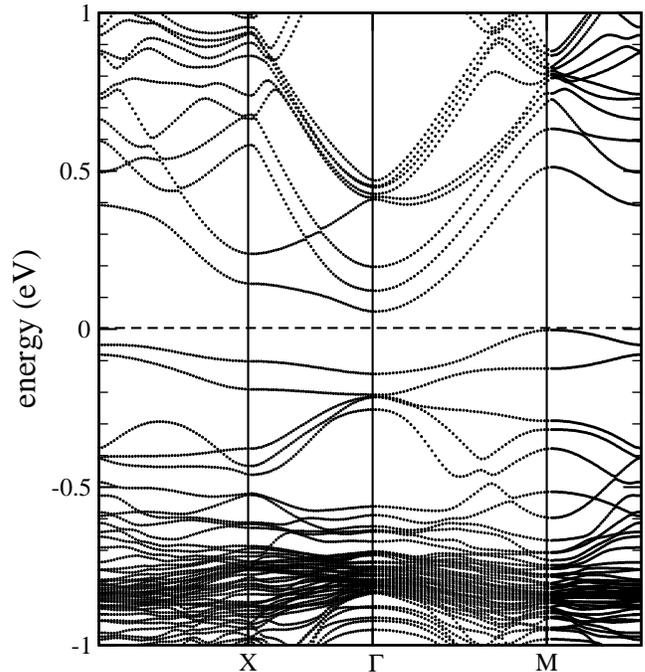}
\caption{Dispersion of the defect state caused by V$_{\rm Ni}$ Ni$_i$ complex in a supercell with 3\% defect concentration.}
\label{fig:bandsZrNiSn-Ni-anti}
\end{figure}

For the calculation of the transport properties and for a better understanding we need the band structure information which is computed with Quantum Espresso.
For each defect type we calculated the band structure of a supercell for the neutral charge state.  
We note that all three defects, at the concentration of 3\% used in our calculation, possess a non-spinpolarized ground state, i.e., our spin-polarized calculations converged to a non-spinpolarized solution. 
Inspection of the calculated Kohn-Sham energy spectra show that the Ni vacancy  gives rise to defect states in the valence band, whereas both the Ni antistructure defect and the additional Ni atom induce in-gap states.
From the supercell band structures plotted in Fig.~\ref{fig:bandsZrNiSn+Ni} and Fig.~\ref{fig:bandsZrNiSn-Ni-anti}, it is clearly seen that the in-gap states form defect-induced bands. The dispersion of the highest occupied impurity band is strongest in the $\Gamma-M$ direction. 
This corresponds to an electronic wavefunction extended along the $\{110\}$ directions, i.e. along the shortest nearest-neighbor distances of the Ni sublattice.
In the supercell, the conduction band minima which had been located at the $X$ points of the primitive unit cell have been folded onto the $\Gamma$ point of the supercell.
For 3\% additional Ni atoms, we observe that the material almost becomes a semi-metal, with the Fermi energy touching the conduction band from below at the $\Gamma$ point and touching the defect band from above at the $M$ point. 
The direct band gap at $\Gamma$ amounts to 0.15~eV. 
This value is in good agreement with experimental reports of a band gap of 0.13~eV from absorption spectroscopy \cite{Schmitt:15} 
and 0.18 eV deduced from electronic transport measurements \cite{Aliev:90} in (supposedly ideal) NiZrSn. 
We believe that this small reported band gap is most likely due to a slight excess of Ni in the samples as a consequence of their fabrication process.
This is in line with the calculated formation energies, since $\rm Ni_i$ is also the defect that forms most easily if excess Ni is available. 
Moreover, it is observed from Fig.~\ref{fig:bandsZrNiSn-Ni-anti} that the highest occupied impurity band in case of the Ni antistructure defect is very similar to the case of additional Ni. 
While the latter had a truly isolated in-gap state, the former one is in touch with the valence band.
We conclude that the defect band is governed by the local environment of the Ni defect, which is very similar in both cases: The Ni defect atom is surround by three or four Ni nearest neighbors, located in the $\{110 \}$ directions. The linear combinations of Ni $3d$ orbitals with bonding character are responsible for the formation of the defect band. This conclusion is confirmed by the inspection of defect wave functions. It is observed that these wave functions have their largest spatial extent in the $\{110\}$ directions. 
This coincides with the direction of the highest dispersion of the defect band.

\begin{figure}[tbh]
%\begin{subfigure}[h]{0.45\textwidth}
%a) \hfill
%
\includegraphics[width=0.48\textwidth,clip]{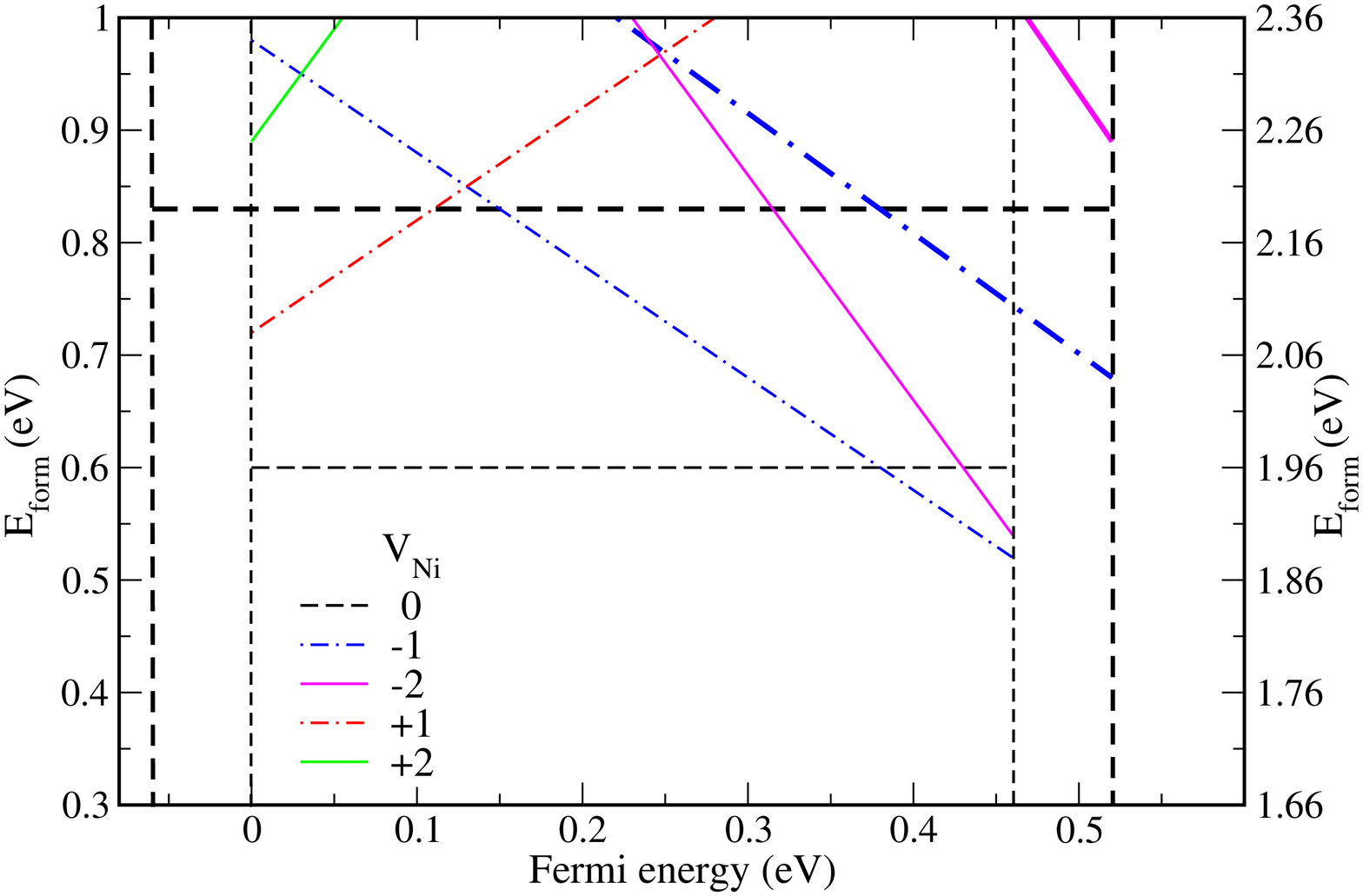}
%\includegraphics[width=0.94\textwidth,clip]{fig8a.eps}
%\caption{}
%\label{fig:form_eng_ni_vac}
%\end{subfigure}
%\begin{subfigure}[h]{0.45\textwidth}

%b) \hfill
%
\includegraphics[width=0.48\textwidth,clip]{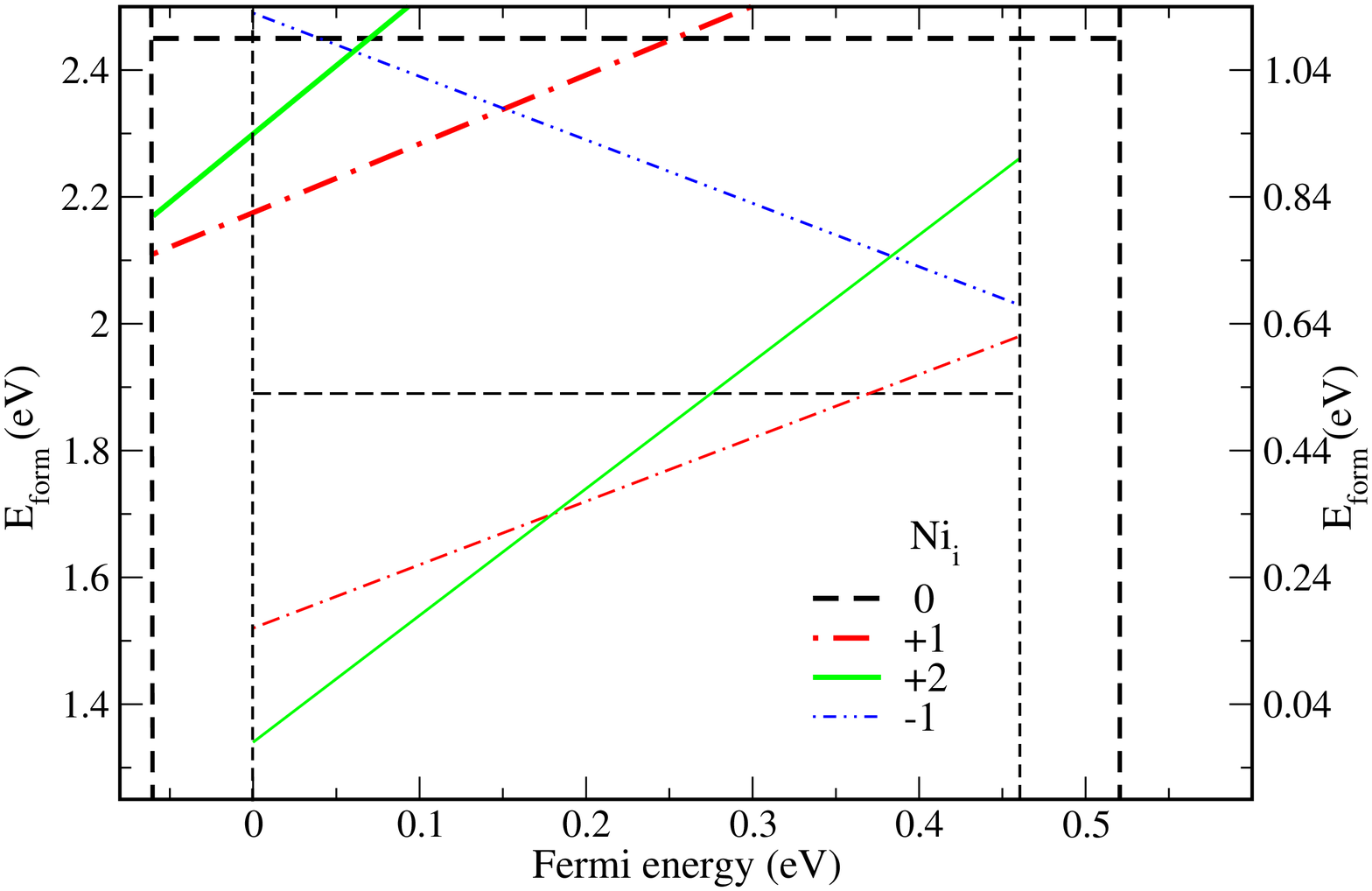}
%\includegraphics[width=0.94\textwidth,clip]{fig8b.eps}
%\caption{}
%\label{fig:form_eng_ni_plus}
%\end{subfigure}
\caption{Formation energies of Ni defects in different charge states for V$_{\rm Ni}$ (upper panel) and Ni$_i$ (lower panel) under Ni-deficient conditions. 
The scale on the left (right) axis gives the formation energy in equilibrium with ZrSn (bulk Ni).
The thick lines indicate results with the HSE06 functional, while the thin lines indicate results with the PBE functional. Vertical lines indicate the position of the valence band and conduction band edges in the supercell. It is assumed that the PBE band gap is located centrally within the (slightly wider) HSE06 band gap.}
\label{fig:charge}
\end{figure}

The formation energies for the different charge states are shown in Fig.~\ref{fig:charge}. The stable charge states of Ni$_i$ are 0, $+1$, and $+2$. 
The charge transfer levels are located at $\epsilon(+2|+1)= 0.18$~eV and $\epsilon(+1|0)= 0.37$~eV.  
In n-type material, the neutral Ni$_i$ can be ionized with an activation energy of 0.09~eV according to the PBE functional. 
In the HSE06 calculation, the neutral charge state of Ni$_i$ persists over the whole upper half of the band gap. 
As the Fermi level is lowered toward the valence band edge, the in-gap state becomes depleted of charge, and the formation energy is considerably  lowered. This explains the stability of the positive charge states. 
The Ni$_i$ atom slightly widens the crystal lattice in its vicinity. Its Ni--Zr or Ni--Sn bonds are about 2\% longer than in bulk, and the volume of the cubic unit cell occupied by Ni$_i$ is enlarged by 6\%.

For V$_{\rm Ni}$, the charge states 0 and $-1$ are possible. In this case, the formation energy is lowered when the Fermi energy approaches the conduction band edge, and the negatively charged vacancy is formed. 
The charge transfer level is located at $\epsilon(0|-1)=0.38$eV.
The volume of the cubic unit cell in which the vacancy is located is contracted by 5\% for the neutral and by about 6\% for the negatively charged vacancy. 

The defect complex V$_{\rm Ni}$Ni$_i$ may occur in four charge states, $-1$,  0, $+1$, and $+2$. The charge transfer levels are 
$\epsilon(0|-1)=0.38$~eV, $\epsilon(+1|0)= 0.28$~eV and $\epsilon(+2|+1)= 0.08$~eV in the PBE calculations.

The overall lowest defect formation energy occurs in the n-type material under Ni-deficient conditions. With $E_F$ at the conduction band edge, the formation energy of V$_{\rm Ni}$ drops to 0.52~eV (0.68~eV using the HSE06 functional).
Even so, the probability of forming vacancies in thermal equilibrium at the growth temperature of 1200~K is expected to be low (less than 1\%).
The formation energy of Ni$_i$ is even higher under equilibrium conditions; however, in the presence of elemental Ni (right scale in Fig.~\ref{fig:charge}b)  
the defect formation becomes possible. 
In annealed samples, additional Ni tends to form Ni$_2$ZrSn precipitates, thereby reducing the concentration of Ni point defects. 
Moreover, intrinsic defects may occur in conjunction with other impurities. Since the Ni$_{i}$ formation energy is lowered in p-type material, doubly positively charged Ni$_{i}$ tends to compensate the extrinsic acceptors in p-type doped samples. This may explain why the attempts to obtain p-type NiZrSn, 
for example by adding V or Sc  as dopants to half-Heusler alloys \cite{Ouardi:10},  
have lead to unsatisfactory results even at large doping concentrations.

Finally, we briefly discuss intrinsic defects of the Co sublattice in CoZrBi. 
To account for a possible magnetic moment at the Co defect, the calculations have been performed with the PBE functional allowing for collinear spin polarization. Only in the supercells containing the Co$_i$ or V$_{\rm Co}$Co$_i$ a spin polarization of 1.55$\mu_{\rm B}$ and $1.18\mu_{\rm B}$, respectively, was found, while the calculations for V$_{\rm Co}$ converged to a non-spin-polarized ground state. 
The defect formation energies for the electrically neutral defects are summarized in Tab.~\ref{tab:Eform}. 
The overall picture is similar to the findings for the Ni defects in NiZrSn.
Note that the formation energies in Tab.~\ref{tab:Eform} are quoted for Co-rich conditions; hence the Co interstitial has a lower formation energy than the Co vacancy. 
The value 1.41~eV for V$_{\rm Co}$ quoted  under Co-rich conditions should be considered an upper bound. 
However, since the full Heusler alloy Co$_2$ZrBi is unstable against decomposition into CoZrBi and elemental Co, the formation energy of Co$_i$ is still positive, even under Co-rich conditions. 
The antistructure defect, V$_{\rm Co}$Co$_{i}$, is again found to have the highest formation energy of the three defect types considered. 
Here, only V$_{\rm Co}$Co$_{i}$ is associated with an in-gap state, while the other two defect types only modify the position of the band edges. 
Similar to Ni$_i$ in NiZrSn, Co$_i$ in CoZrBi forms with less energy cost in its positive charge states,  
and may thus act in compensation of extrinsic dopants.  
The formation energy in p-type material, where $E_F$ lies at the valence band edge, is calculated to be 0.75~eV using the PBE functional.
This value, which is the lowest one we observed for all Co defects considered, is significantly higher than the formation energies calculated in NiZrSn. 
We thus believe that intrinsic Co defects are rather unimportant and are created only in negligible concentrations, at least in thermal equilibrium.

\subsection{Thermoelectric properties \label{sec:Seebeck} }

Finally, we address the thermoelectric properties of the two half-Heusler compounds studied here, including the effect of the most probable intrinsic defects. 
Using semiclassical Boltzmann transport theory within the constant relaxation time approximation, one can show that the band structure information {\em alone} is sufficient to calculate the Seebeck coefficient, since it is given by a quotient in which the relaxation time cancels out. 
The results below are based on the PBE band structure. 
As a highly resolved (in the Brillouin zone) representation of the electronic bands is needed, we further refine the Kohn-Sham band structure obtained from the DFT calculations on a dense mesh by interpolation.\cite{Pickett:88}  
This interpolation procedure is implemented in the  Boltztrap code \cite{Madsen:06} that was employed to calculate the electrical transport properties. 

\begin{figure}[bth]
		\includegraphics[width=0.45\textwidth]{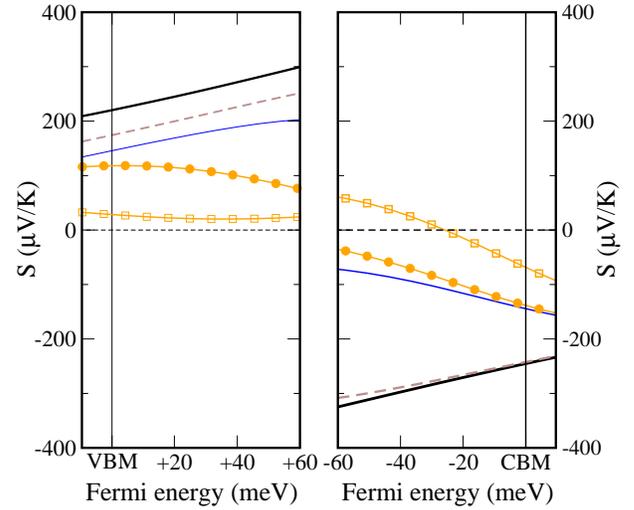}
\caption{Seebeck coefficient at $T=600$ K near the band edges for pure NiZrSn and for samples containing defects of one type at the 3\% concentration level: Bulk NiZrSn (thick black), Ni vacancy (brown, dashed), interstitial Ni (thin blue), and V$_{\rm Ni}$Ni$_i$ antistructure complexes (orange, with symbols). For the latter, the transport directions  perpendicular to the displacement vector $\vec d$ of the defect complex (orange circles) and parallel to $\vec d$ (orange square) must be distinguished. }
\label{fig:seebsplit}
\end{figure}

Presenting our results in Fig.~\ref{fig:seebsplit}, we treat the Fermi energy as a variable because it  strongly  depends on doping.  
While specific choices for donors or acceptors could be worth a separate study, 
we prefer to investigate the generic case, and thus cover a range of shallow donors or acceptors by presenting the Seebeck coefficient in energy  intervals of 60~meV above or below the band edges, respectively. 
Fig.~\ref{fig:seebsplit} compares NiZrSn with various types of intrinsic defects to bulk NiZrSn. The latter, defect free material is found to have the highest thermopower both in the p-doped and the n-doped regime. 
For the ease of comparison, we have aligned the VBM of all samples in Fig.~\ref{fig:seebsplit}a) and the CBM of all samples in Fig.~\ref{fig:seebsplit}b). 
The Fermi energy is given relative to the respective band edge. 
The defect-induced changes of the electronic structure lead in all cases to a reduction of the absolute magnitude of the Seebeck coefficient. 
For samples with V$_{\rm Ni}$, we observe only small changes of the Seebeck coefficient near the band edges. 
At the VBM, the curve $S(E_F)$ has a nearly constant offset of about 50 $\mu$V/K compared to bulk, while at the CBM it coincides with the bulk curve. 
The more pronounced reduction of the thermopower for samples containing  either Ni$_i$ or V$_{\rm Ni}$Ni$_{i}$ is due to 
the in-gap state induced by these defects. 
In the case of Ni$_i$, the defect state is separated from both the valence and conduction bands, while in the case of V$_{\rm Ni}$Ni$_i$ it is separated from the conduction band only. 
This is why the behavior of $S(E_F)$ in case of V$_{\rm Ni}$ (thin blue curve in Fig.~\ref{fig:seebsplit}) is nearly parallel to the other two, while V$_{\rm Ni}$Ni$_i$ (orange curves, with symbols) shows  different behavior.  
In the p-doped case with antistructure defects, the Seebeck coefficient is small because the Fermi energy lies in a region with a finite density of states due to the hybridization of valence and in-gap states that fill the whole lower half of the former band gap (cf. Fig.~\ref{fig:bandsZrNiSn-Ni-anti}). 
Moreover, the Seebeck coefficient is anisotropic, since the displacement vector $\vec d$ of the Ni atom breaks the symmetry.
However, 
since realistic samples will have V$_{\rm Ni}$Ni$_i$ defects with Ni atoms displaced arbitrarily in all directions of the $\{110\}$ family, the anisotropy is expected to average out in measurements on macroscopic samples.  
In n-doped samples, (cf. Fig.~\ref{fig:seebsplit} b) the Seebeck coefficient in the sample with the V$_{\rm Ni}$Ni$_i$ defects even changes its sign in the direction parallel to $\vec d$ depending on the position of the Fermi energy. In the direction perpendicular to $\vec d$, it remains negative but is strongly reduced in magnitude. The latter also holds true for samples with Ni$_i$ defects. 
This reduction can be attributed to a compensation effect of electrons at the CBM and hole carriers in 
the in-gap state both contributing with opposite sign to $S(E_F)$. From our band structure and effective mass data we conclude that the mobility of the two carrier types is very different; large for the electrons and small for the holes, in particular in the more realistic situation of randomly distributed defects. 
This finding is in line with an earlier interpretation of ambipolar conduction observed in p-doped ZrNiSn that was attributed to widely different carrier mobilities.\cite{Schmitt:15} 

Experimentally, the Seebeck coefficient in n-doped NiZrSn has been studied by several groups. 
Measurements at room temperature  in well-annealed samples \cite{Qiu:10} found values as large as $-345 \mu$V/K, with the absolute values decreasing at higher $T$. 
The latter behavior is indicative of semiconducting samples, demonstrating that the defect concentration was sufficiently low for a gap to open up right below the CBM.   
The calculated maximum thermopower that we obtained for bulk NiZrSn is about $-400\mu$V/K if the Fermi energy is located in the gap.  
Since the precise position of the Fermi energy in the experimental samples is not known, the agreement between experiment and our calculation must be considered satisfactory. 
However, many samples used in experiment show indications of in-gap states, reflected by a smaller thermopower and a quasi-metallic behavior as function of temperature.  For example, Zou {\em et al.} report a Seebeck coefficient of  $-110 \mu$V/K at room temperature that changes to $-180 \mu$V/K at 700~K \cite{Zou:13}. 
Earlier work \cite{Uher:99} reported a very small Seebeck coefficient below 150~K, rising to $-210 \mu$V/K at room temperature.  
This could be seen as in indication that $E_F$ was initially lying near a defect band at low temperature and only later rises into the gap between defect band and CBM.
According to our calculations, such samples may contain Ni$_i$ or V$_{\rm Ni}$V$_i$ defects. The values of around $-140 \mu$V/K in defective samples at 600~K which we find for $E_F$ close to the CBM agree well with the experimentally observed numbers.

\begin{figure}[bth]
		\includegraphics[width=0.5\textwidth]{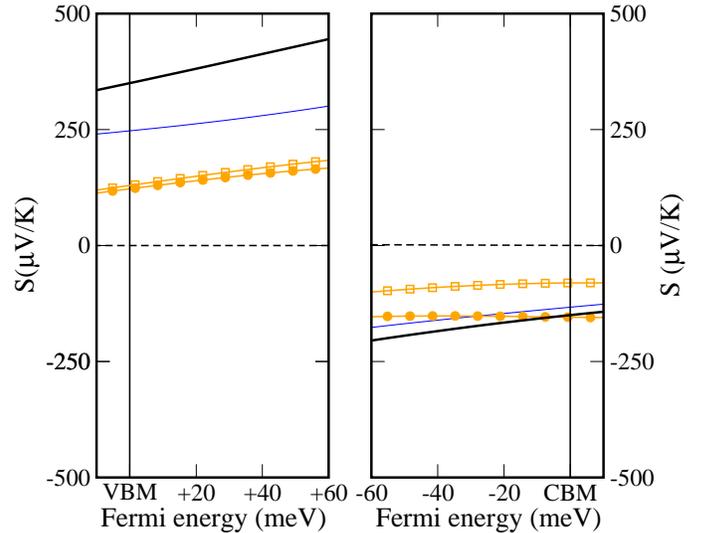}
\caption{Seebeck coefficient at $T=600$ K near the band edges for pure CoZrBi and for samples containing defects of one type at the 3\% concentration level: Bulk CoZrBi (thick black),  interstitial Co (thin blue), and V$_{\rm Co}$Co$_i$ antistructure complexes (orange, with symbols). For the latter, the transport directions  perpendicular to the displacement vector $\vec d$ of the defect complex (orange circles) and parallel to $\vec d$ (orange square) must be distinguished. }
\label{fig:seebsplitCo}
\end{figure}

The alternative material CoZrBi shows thermoelectric behavior qualitatively similar (see Fig.~\ref{fig:seebsplitCo})  to NiZrSn. However, under p-type doping, ZrCoBi displays a larger thermopower than NiZrSn, while the opposite holds for n-type samples. 
The large thermopower for p-type CoZrBi can be traced back to the rather flat valence band along the $\Gamma - L$ direction in the band structure plot (cf. Fig.~\ref{fig:bands-ZrCoBi} ).
The valence band maxima at the $L$ points are energetically close to the VBM at $\Gamma$ and hence additionally contribute to the thermopower of the hole carriers. 
As we have already seen for NiZrSn, the sample with the ideal crystal structure (without defects) is expected to yield the largest thermopower also in case of CoZrBi.
For samples containing Co$_i$ in appreciable concentration, the thermopower is reduced, but still follows the trend of the defect-free sample. 
In samples with V$_{\rm Co}$Co$_i$ defect complexes, the thermopower is found to be strongly reduced.  Again, as in NiZrSn, this reduction, which is particularly strong under n-type conditions, can be attributed to a compensation effect between hole carriers in the in-gap state and electron-like carriers in the conduction band. 
Unfortunately we are not aware of any measurements of the Seebeck coefficient in CoZrBi which our results could be compared to. 
However, in view of the rather large thermopower in defect-free CoZrBi under p-type conditions, it may be worth while studying experimentally the thermoelectric properties of this material.

The applicability of NiZrSn and CoZrBi for thermoelectric applications depends not only on the thermopower, but also on the electrical conductivity $\sigma$ and thermal conductivity $\kappa$ of the samples. 
The thermoelectric power factor $\sigma S^2$ requires for its calculation knowledge of the transport relaxation time $\tau$ which is not generally available.
Using the formalism for a constant, energy-independent relaxation time one obtains values of
$\sigma S^2 = 0.0047 \, {\rm W/(m \cdot K}^2) \tau/10^{14}$s for $E_F = E_{\rm VB}$ and $\sigma S^2 = 0.003 \, ({\rm W/m \cdot K}^2)  \tau/10^{14}$s for $E_F = E_{\rm CB}$ in ideal NiZrSn at $T=600$~K. Typical values of $\tau$ are expected to be in the range of 10$^{14}$~s. 
For ideal CoZrBi, the corresponding values are $\sigma S^2 = 0.011 \, ({\rm W/m \cdot K}^2) \tau/10^{14}$s for $E_F = E_{\rm VB}$ and $\sigma S^2 = 0.007 \, ({\rm W/ m \cdot K}^2)  \tau/10^{14}$s for $E_F = E_{\rm CB}$.  
Hence, CoZrBi may a promising material for the p-leg of thermoelectric devices.

The lattice thermal conductivity $\kappa$ of ideal samples is known from calculations \cite{Carrete:14x} to be close to 20~W/(K m) at $T = 300$~K.   
Similar values were obtained from previous qualitative estimates by us.\cite{Fiedler:15b}   
However, in samples that are strongly doped and/or contain a high concentration of intrinsic defects, these provide additional sources for scattering, and both the mobility and the thermal conductivity are considerably reduced compared to ideal samples. 
Experimental values for NiZrSn display considerable spread, probably due to varying structural quality of the samples.
At room temperature, measured values of 12.5~W/(K  m) (Ref. \onlinecite{Qiu:10}), 17.2~W/(K m) (Ref.~\onlinecite{Uher:99}) and 9.5~W/(K m) (Ref.~\cite{Zou:13})  have been reported, all lower than the theoretical results. One reason for this desirable reduction  could be phonon scattering from structural imperfections in real samples. 

\section{Conclusion}
In summary, we have studied the electronic band structure of the ternary compound semiconductors NiZrSn and CoZrBi.
Both materials have their conduction band minima at the X point of the Brillouin zone and possess highly anisotropic effective masses. 
The shear deformation potentials $\Xi_u$ are unusually low, and hence pure samples of these materials are expected to show high mobilities in the n-type transport regime.
Both materials have valence band maxima at the $\Gamma$ point, with CoZrBi having additional maxima at the $L$ points. 
The latter finding is responsible for the predicted high thermopower of CoZrBi under p-type conditions.
In the phononic band structure, it is found that the optical modes overlap with the acoustic modes in ZrNiSn, whereas CoZrBi  
displays a gap in its phononic density of states between optical and acoustic branches. 
The experimentally observed ineffectiveness of p-doping in NiZrSn could be rationalized by the spontaneous formation of interstitial Ni defects which, in their positive charge states, counteract the desired p-doping. 
On the other hand, intrinsic point defects pose no obstacle to  n-doping in NiZrSn,
The Co-related point defects in CoZrBi have a higher formation energy than their counterparts in NiZrSn, and hence appear to be less important.
While interstitial Ni could be exploited as additional donors in n-type NiZrSn, and could thus enhance the electrical conductivity, all 
point defects are found to diminish the thermopower compared to the ideal crystals. 
In-gap states are induced by Ni$_i$ and V$_{\rm Ni}$Ni$_i$ in NiZrSn, and by V$_{\rm Co}$Co$_i$ in CoZrBi. 
In the latter material, both the  Co$_i$  and the V$_{\rm Co}$Co$_i$  defect are predicted to display a magnetic moment, and could thus be accessible to electron spin resonance studies.
While NiZrSn has already been synthesized and its thermoelectric properties studied, CoZrBi samples have not been characterized in detail so far. 
The predicted high thermopower in the p-type regime could make this material an interesting candidate to be used in the p-leg of thermoelectric generators. 

\section*{Appendix}
Elastic constants are calculated by putting strain on a cubic unit cell of the materials. 
In cubic, elastically isotropic media, the elasticity tensor contains only three independent elements: $C_{11}$, $C_{12}$ and $C_{44}$. $C_{11}$ can be calculated by the change in DFT total energy $E(x)$ when the unit cell is strained in one direction (uniaxial strain). With this data and the relation 
\begin{equation}
C_{12}=\left(2\frac{d^2}{dx^2}E(x)/3V - C_{11}\right)/2 \, ,
\end{equation}
using $x= \epsilon_{11}=\epsilon_{22}=\epsilon_{33}$, $C_{12}$ can be obtained. $C_{44}$ becomes important when shear strain is applied along the (111) direction under volume conservation. The strain tensor for this case is  
\begin{equation}
{\bf \epsilon} = 
\begin{pmatrix}
 0 &x &0 \\ 
x &0 & 0\\
0 & 0& \frac{x^2}{1-x^2}\\
\end{pmatrix}
\end{equation}
Again, $C_{44}$ is obtained from the second derivative of the DFT total energy $E(x)$. 

For electron-phonon interaction models, the deformation potentials for 
the valence band maximum (VBM) and the conduction band minimum (CBM) are needed. 
The type of the deformation potential depends on the symmetry. In the case of CoZrBi and NiZrSn, the VBM at the $\Gamma$ point is three-fold degenerate. 
Hydrostatic pressure only shifts the position in energy, while shear strain leads to a splitting of the VBM degeneracy. 
Three deformation potentials $a_v$ (hydrostatic) and $b$, $d$ (shear) are sufficient to determine strain effects on the VBM.

The CBM is located at the  $X$ point in reciprocal space at the zone boundary. 
For this case, which is similar to the conduction band in silicon, two deformation potential are sufficient, labelled $\Xi_u$ and $\Xi_d$ in the notation of Herring and Voigt \cite{Herring:56}. 
$\Xi_d$ describes the coupling to the trace of the strain tensor, while $\Xi_u$ describes coupling to uniaxial strain in the direction $\Gamma X$.   
$\Xi_u$ is a shear deformation potential and $a_c = \Xi_d+ {1 \over 3} \Xi_u$ represents the volume deformation. 
The two volume deformation potentials $a_v$ and $a_c$ are defined by
\begin{equation}
%$$
\Delta E_{v/c} = a_{v/c} \frac{\Delta V}{V} \, ,
%$$
\end{equation}
where $\Delta E_{v/c}$ is the change of the valence band and conduction band edge, respectively, and $\Delta V$ is the change in the unit cell volume $V$. 
Both $a_v$ and $a_c$ is obtained from inspection of the Kohn-Sham eigenvalues in DFT calculations with varying lattice constant. 
To obtain $b$, $d$, and $\Xi_u$, we perform DFT calculations for unit cells subject to uniaxial deformation or shear strain, analogous to the ones used to determine the elastic constants $C_{12}$ and $C_{44}$. For each deformation potential, a specific set of deformed unit cells was considered. Then, the changes in the Kohn-Sham eigenvalues are fitted to the behavior expected from deformation potential models, as found in Ref.~\onlinecite{vandeWalle:89}  and  \onlinecite{Christensen:84}. 

\section*{Acknowledgments}
The authors gratefully achnowledge financial support from the Priority Program SPP1386 
of Deutsche Forschungsgemeinschaft (DFG) under grant number KR2057/4-2.

\end{document}